\documentclass[%
 reprint,
 amsmath,amssymb,
 aps,
 longbibliography
]{revtex4-1}

\usepackage{graphicx}
\usepackage{dcolumn}
\usepackage{bm}
\usepackage{caption}
\usepackage{subcaption}
\usepackage{gensymb}
\usepackage{xcolor}

\newcommand{\bfp}{{\bf p}}
\newcommand{\bfx}{{\bf x}}
\newcommand{\bfn}{{\bf n}}
\newcommand{\bfe}{{\bf e}}
\newcommand{\bfv}{{\bf v}}
\newcommand{\bfc}{{\bf c}}
\newcommand{\bfF}{{\bf F}}

\newcommand{\bfgam}{{\bf \gamma}}

\begin{document}

\preprint{APS/123-QED}

 \title{A Minimal Model of the Hydrodynamical Coupling of Flagella on a Spherical Body with application to Volvox}

\author{Forest O. Mannan}
\email{fmannan@western.edu}
\affiliation{%
 Mathematics \& Computer Science Department, Western Colorado University,
1 Western Way
Gunnison, CO 81231}
 
 \author{Miika Jarvela}
\author{Karin Leiderman}%
 \email{kleiderman@mines.edu}
\affiliation{%
Department of Applied Mathematics and Statistics, Colorado School of Mines, 1500 Illinois St., Golden, CO 80401
}

\date{\today}

\begin{abstract}

Flagella are hair-like appendages attached to microorganisms that allow the organisms to traverse their fluid environment. The algae \textit{Volvox} are spherical swimmers with thousands of individual flagella on their surface and their coordination is not fully understood. In this work, a previously developed minimal model of flagella synchronization is extended to the outer surface of a sphere submerged in a fluid.  Each beating flagellum tip is modelled as a small sphere, elastically bound to a circular orbit just above the spherical surface and a regularized image system for Stokes flow outside of a sphere is used to enforce the no-slip condition. Biologically relevant distributions of rotors results in a rapidly developing and robust symplectic metachronal wave traveling from the anterior to the posterior of the spherical \textit{Volvox} body.

\end{abstract}

\pacs{Valid PACS appear here}

\maketitle

Cilia and flagella are ubiquitous among eukaryotic cells. These small, hair-like appendages extend from cell membranes and play important roles in locomotion and fluid transport by undergoing a periodic motion. Examples include the transport of foreign particles out of the lungs \cite{TWSC14}, the creation of left-right asymmetry in embryonic development \cite{essner2002left}, and filter feeding \cite{mayne2017particle}.

These biologically relevant flows are generally created through the coordinated collective motion of many cilia or flagella. The origin and means of this large-scale coordination has been a long standing area of research \cite{taylor1951analysis}. In some scenarios, hydrodynamic coupling alone has successfully explained such coordination \cite{brumley2014flagellar,brumley2015metachronal, vilfan2006hydrodynamic, goldstein2016elastohydrodynamic, dillon2000integrative, yang2008integrative, guo2018bistability}. Experimental approaches range from investigating colloidal oscillators with optical tweezers \cite{Kotar7669} to observing synchronization between lone flagellum pairs, emanating from two separate cells and tethered at fixed distances via  micro-pipettes \cite{brumley2014flagellar}. Examples of theoretical approaches include the study of filaments with internal driving forces immersed in a fluid \cite{mannan2018, dillon2000integrative, goldstein2016elastohydrodynamic, yang2008integrative, guo2018bistability} and so-called minimal models where the cilia or flagella are represented as oscillating `rotors' immersed in a viscous fluid \cite{niedermayer2008synchronization, brumley2015metachronal, brumley2016long}. This latter approach is what we build on in the current work.  

Ensembles of large numbers of cilia often exhibit regular variations in the beating phase of adjacent cilia, which are characterized as metachronal waves (MWs) \cite{elgeti2013emergence, brumley2015metachronal,mitran2007metachronal}. The colonial alga \textit{Volvox carteri} (Volvox) has become a model organism for studying the emergence of MWs \cite{brumley2015metachronal,matt2016volvox}; an informative review of these studies can be found elsewhere \cite{goldstein2015green}. \textit{Volvox} is a multicellular green algae whose surface consists of fairly regularly-spaced biflagellated somatic cells, embedded in the extracellular matrix \cite{matt2016volvox, kirk2005volvox}. \textit{Volvox} swimming is mainly due to the coordinated beating of their flagella, which exhibit clear MWs traveling from the anterior to posterior of the spherical \textit{Volvox} body \cite{brumley2015metachronal,brumley2012hydrodynamic}. Further, \textit{Volvox} flagella beat towards the posterior of the colony with a small 10-20 degree tilt out of the meridional plane \cite{hoops1983ultrastructure,hoops1997motility}.  The tilt has long been thought to allow volvox to `swirl', where they rotate during forward progression swimming \cite{mast1926reactions,hoops1997motility}. 

Minimal models of coupled rotors \cite{niedermayer2008synchronization,brumley2014flagellar} are particularly amenable to the theoretical study of MW formation on \textit{Volvox} due to the  number and spacing of flagella on the \textit{Volvox} surface; one flagellum is close enough to another flagellum to influence its periodic beating (via hydrodynamics) but typically not close enough to make physical contact. To represent a single flagellum with a rotor, the tip of the flagellum is modeled as a small, rigid sphere with a preferred circular orbit. The shape of the orbit is controlled with a system of springs and the motion is due to a prescribed driving force. The fluid flow induced by one rotor on another rotor can then be well approximated by a single Stokeslet \cite{brumley2014flagellar}. Additionally, the leading-order far-field flow induced by a rigid sphere is precisely given by a Stokeslet \cite{nasouri2016hydrodynamic}. Thus, a model rotor (oscillator) captures both the phase of the beating flagellum and well approximates its corresponding, induced far-field flow. 

Previous studies of \textit{Volvox} flagella with minimal models of coupled oscillators were able to reproduce semi-quantitative characteristics of the average metachronal dynamics and the emergence of MWs \cite{brumley2015metachronal,brumley2012hydrodynamic}, while also using simplifying assumptions about \textit{Volvox} geometry, e.g., the surface of \textit{Volvox} was treated as a no-slip plane. One study considered flagellum beating on a spherical body, but was limited by using a single chain of rotors that all beat in the same direction \cite{nasouri2016hydrodynamic}; flagella on \textit{Volvox} cover the entire surface and beat from the anterior to the posterior. In this study, we extend these minimal models of coupled oscillators to investigate biologically-relevant distributions of beating flagella on the surface of a sphere.

\begin{figure}
    \centering
      \begin{subfigure}[t]{0.49\textwidth}
 \includegraphics[width=\textwidth, trim={0cm 0cm 0cm 0cm}]{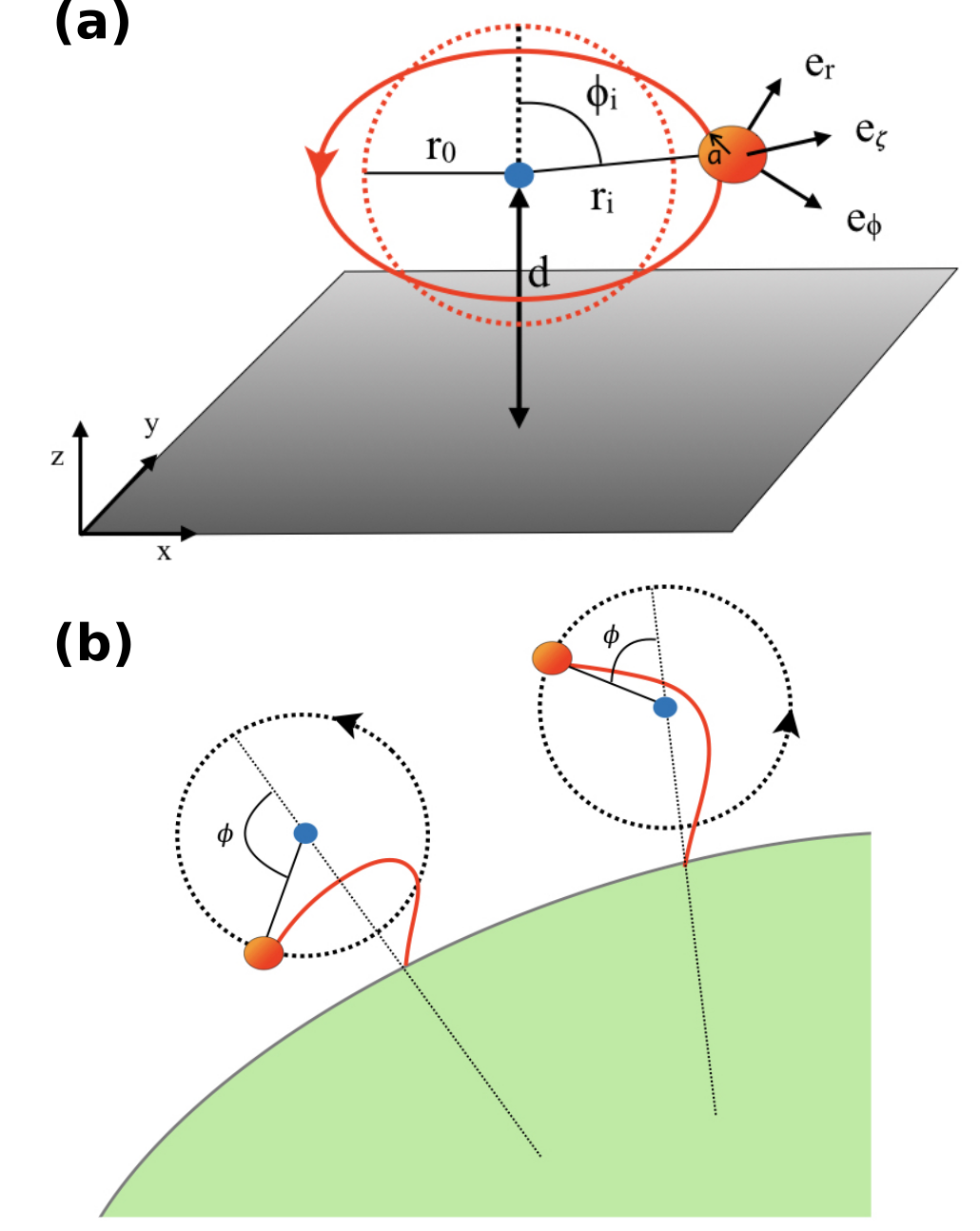}
 \end{subfigure}

	\caption{(a) A schematic of the rotor model of the \textit{Volvox} flagellum in which a small rigid sphere is elastically bound to a preferred trajectory; adapted from previous work \cite{brumley2015metachronal,brumley2012hydrodynamic}. (b) The rotor position reflects different configurations of the flagellum}
	\label{fig:RotorSchematics}
\end{figure}

 Following the studies by Brumley \textit{et al. } \cite{brumley2015metachronal,brumley2012hydrodynamic}, each rotor is a rigid sphere of radius $a$, elastically bound in a circular trajectory of radius $r_0$, about a prescribed center point located a distance $d$ above the spherical \textit{Volvox} body, as depicted in Figure \ref{fig:RotorSchematics}(a). The preferred plane of orbit is defined by the center of rotation and a vector normal to the plane of rotation, $\bfn$. The orbit of the rotor is driven by a constant tangential driving force $f^{dr}$ in the $\bfe_\phi$ direction and the preferred trajectory is elastically enforced through a radial spring and a transverse spring normal to the plane.
 
To evolve the positions of the rotors in time, the velocity of each rotor is determined from a system of coupled, force-balance equations, one for each rotor. The forces acting on each rotor are the elastic spring forces that resist stretching, the net hydrodynamic drag force, and the prescribed constant driving force. 

In the case of one single rotor, the hydrodynamic drag force is assumed to be equal and opposite to the driving force and spring forces yielding the force balance:
 \begin{equation}\label{eq:SingleRotorForceBalance}
     \bfgam(\bfx)\bfv = -\lambda(r-r_0)\bfe_r - \eta \zeta \bfe_\zeta + f^{dr}\bfe_\phi,
 \end{equation}
\noindent where $\lambda$ and $\eta$ prescribe the stiffness of the radial and transverse springs and $\bfgam$ is the friction tensor. For simplicity, as in previous studies, we let $\bfgam = \gamma_0 {\bf I}$ where $\gamma_0 = 6\pi\mu a$, the drag on a sphere in free space, and $\mu$ is the dynamic viscosity of the fluid \cite{nasouri2016hydrodynamic}. With the parameters used in this study we compared this free-space drag to the case if the rotor was above the actual spherical body, and estimated a relative difference of about $2.7\%$, see the Supplementary Material for details \cite{SupplementaryMaterial}.

\begin{figure*}
    \centering
    \begin{subfigure}[t]{\textwidth}
\includegraphics[width=\textwidth]{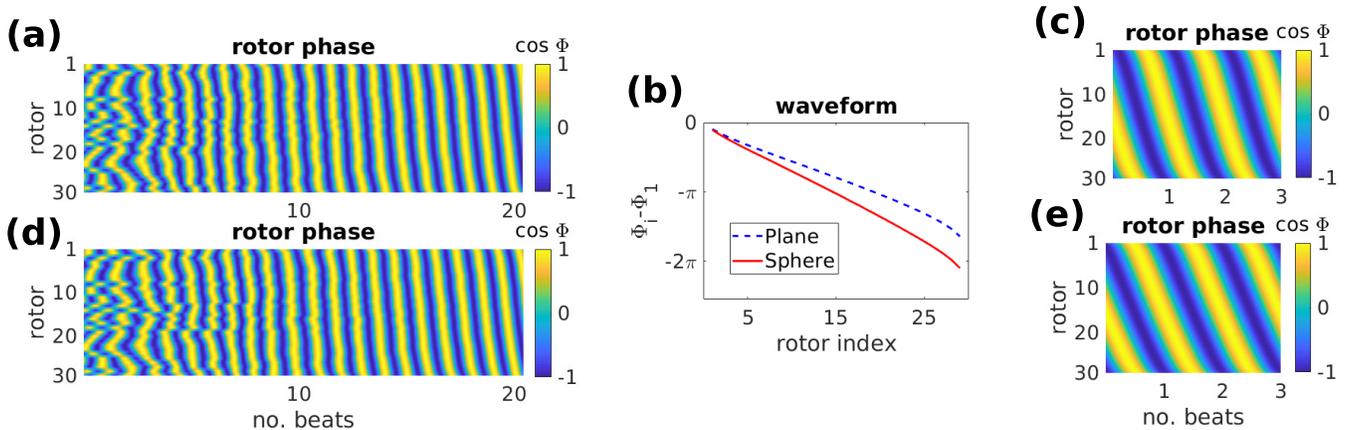}
\end{subfigure}
	\caption{The evolution of a chain of 30 rotors above a plane and a chain of rotors extending halfway around a sphere are considered. The same random initial relative configuration is considered in either case. (a) shows the evolution of the phase over the first 20 periods and (c) shows the phase after 1200 beats, for the case of a plane while (d) and (e) is for the case of a sphere. (b) shows the final waveform for both a sphere and a plane after 1200 beats. These respective steady states are reached regardless of the initial rotor positions.}
	\label{fig:SingleChainDynamics}
\end{figure*}

When considering a single lone rotor, there is no imposed external fluid flow and thus the hydrodynamic drag on the rotor depends only on the rotor's own velocity. To evolve $N$ coupled rotors in time, a net drag force on each individual rotor must be considered that includes the effects of the external flow induced by all the other rotors. The external fluid flow imposed on a single rotor by all other rotors is calculated using a far-field approximation with regularized Stokeslets \cite{cortez2001method,cortez2005method}. Letting $G$ be the regularized Green's function in the presence of a no-slip sphere \cite{wrobel2016regularized}, $\{\bfx_i \}_{i=1}^N$ be the rotor locations, and $\bfF^\text{ext}_j$ be the external forces acting on the $j^\text{th}$ rotor then the net hydrodynamic force on the $i^\text{th}$ rotor is ${\bf F_i} = -\bfgam(\bfx_i)[\bfv_i-\sum_{j\ne i} G(\bfx_i,\bfx_j)\bfF_j^\text{ext}]$. As such, the force balance on the $i^\text{th}$ rotor is given by
 \begin{align}\label{eq:CoupledForceBalance}
 -{\bf F_i} = &-\lambda(r_i-r_0)\bfe_r- \eta \zeta_i \bfe_\zeta + f^{dr}\bfe_\phi,
 \end{align}
\noindent where $r_i = \|\bfp_i-\bfc_i \|$ and $\bfp_i$ is the projection of the $i^\text{th}$ rotor's location onto its respective preferred plane of orbit and $\zeta_i = \|\bfp_i - \bfx_i\|$ is the distance from the $i^\text{th}$ rotor to its preferred plane of orbit. This gives rise to a $3N\times 3N$ system of linear equations for the rotor velocities. We note that the free-space drag assumption results in $\bfgam$ having a strictly diagonal form,  which allows for efficient calculation of the unknown fluid velocities in Eq.~\ref{eq:CoupledForceBalance}.

For the regularized Green's function, $G$, the regularization parameter $\epsilon$ is chosen to be equal to $a/d$ and the blob function
\begin{equation*}
    \psi_\epsilon(r) = \frac{15\epsilon}{8\pi(r^2 + \epsilon^2)^{7/2}},
\end{equation*}
\noindent is used. The sensitivity of the results to this choice of regularization parameter was investigated by considering its effect on the phase differences between two rotors above a plane. As discussed in the Supplementary Material \cite{SupplementaryMaterial}, variation in the regularization parameter led to negligible effects on the dynamics.

To study MW formation outside a sphere with this model, we chose one set of parameters close to those from previous studies and that are reflective of \textit{Volvox} flagella  \cite{niedermayer2008synchronization, brumley2012hydrodynamic,brumley2015metachronal}. We set $d=$ 10 $\mu$m, $r_0 =$ 5 $\mu$m, and $a=$ 1 $\mu$m, which approximates a flagellum that is about 1 $\mu$m thick and is about 15 $\mu$m long when fully extended, refer again to Figure \ref{fig:RotorSchematics}(a). The driving force is $f^\text{dr} = 2\pi r_0 \gamma_0 / T$ where $T = 1/33$ s, to give an approximate beat frequency of 33 Hz \cite{brumley2012hydrodynamic}.  We also set the dimensionless spring stiffness ratios to be $\Lambda = \lambda d/ f^\text{dr} = 0.1$ and $\eta d / f^\text{dr}=0.1$ \cite{niedermayer2008synchronization, brumley2012hydrodynamic,brumley2015metachronal}. A \textit{Volvox} radius of $200 \mu m$ is considered. These parameters are used for all the simulations presented in this work.

The phase of a given rotor, $\Phi(\phi)$, is computed by post processing the dynamic rotor positions such that $\dot\Phi$ is constant during a given period. In turn, a period of a given rotor is defined as the time that elapses as $\phi$ ranges from 0 to $2\pi$ where $\phi$ is calculated from the projection of the rotor onto its preferred plane of orbit. It should be noted that using the regularized Green's function in the presence of a no-slip sphere results in each rotor pushing a larger net volume of fluid at the apex of each orbit than at the nadir. This naturally mimics the power and recovery stokes, respectively, of flagella. As such, the $\phi$ range of $[3\pi/2,2\pi)\cup[0,\pi/2)$ and $[\pi/2,3\pi/2)$ can be thought of as corresponding to power and recovery strokes respectively.

For each simulation in this study, the desired spatial distribution of rotor centers was chosen and each was assigned a random initial phase. Next, for each rotor, the following steps were repeated until a final desired time: the rotor velocity, ${\bf v}$ was determined from the full system of coupled force-balance equations, then the rotor position, ${\bf x}$, was updated by numerically integrating $d{\bf x}/dt = {\bf v}$ with a second-order Runge-Kutta method.

We first studied the evolution of a single chain of 30 rotors, equally spaced along half of a sphere of radius 200 $\mu$m, with randomly chosen initial phases. The arclength between the centers of adjacent rotor's trajectories was set to $2d$, to mimic the chain of rotors studies above a plane in previous studies \cite{brumley2012hydrodynamic,brumley2015metachronal}, except with curvature from the spherical surface. No matter the initial phases of the rotors, a single steady state was always achieved. The resulting phases from each simulation decreased from one rotor to its neighbor in the anterior-posterior direction, and thus the steady states observed are symplectic metachronal waves as previously observed in \textit{Volvox} \cite{brumley2012hydrodynamic,brumley2015metachronal}. These results are in line with the parameter choices as they were stated to reside in the symplectic MW regime \cite{brumley2012hydrodynamic}. To directly compare our results to previous ones, the simulation was repeated with a chain of 30 rotors above a plane, whose centers are a distance $2d$ apart, and using the image system for regularized Stokeslets above a plane \cite{ainley2008method}. Figure~\ref{fig:SingleChainDynamics} compares these two cases with the exact same random initial configurations, relative to the rotor positions in the chain. The phase profile for the case above a plane compares well to that of previous studies (compare our Figure~\ref{fig:SingleChainDynamics}(b) to Figure~3 for $\Lambda=0.1$ in \cite{brumley2012hydrodynamic}). The phases in both the plane and sphere case evolve similarly though the final waveform in the sphere case exhibits a greater total variation in phase difference, i.e., on the surface of a sphere, there is a slightly greater phase difference between each adjacent rotors.

\begin{figure*}
    \centering
    \begin{subfigure}[t]{0.99\textwidth}
        \includegraphics[width=\textwidth]{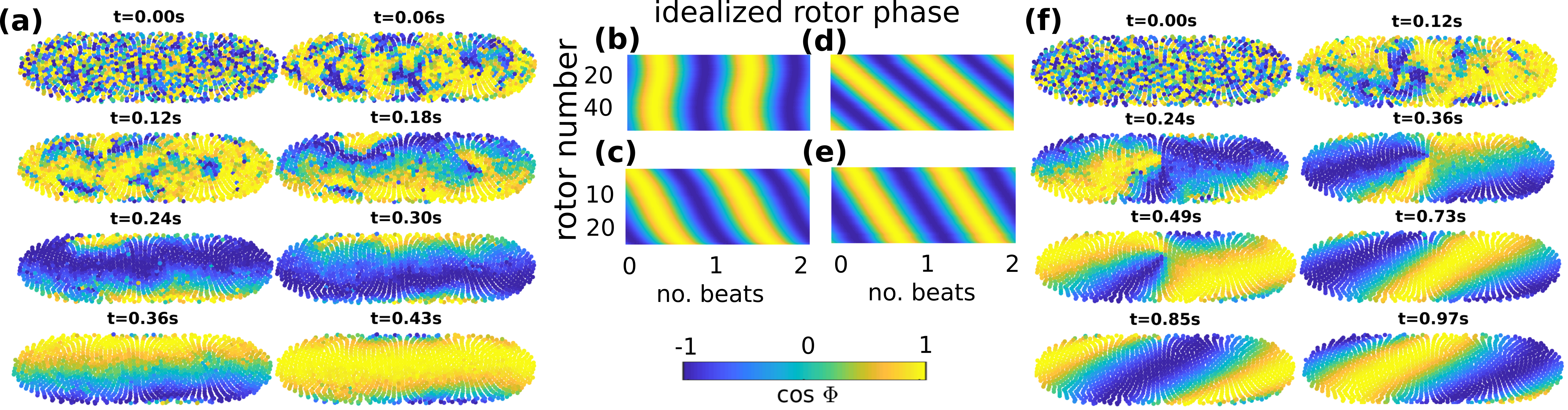}
    \end{subfigure}
	\caption{Snapshots of the evolution of the phase of an approximately uniform distribution of 1257 rotors beating without (a) and with tilt (f), using a type IV Eckert projection starting from random initial phases. The time between snapshots correspond to two and four periods of a lone rotor for (a) and (f) respectively. The phase along a chain of idealized rotors around the equator of the \textit{Volvox} without (b) and with tilt (d). The phase of along a chain of idealized rotors in  the meridional direction from the anterior to the posterior without (c) and with tilt (e).}
	\label{fig:UniformDynamics}
\end{figure*}

In this study, we will assume that the somatic cells on the surface of \textit{Volvox} are roughly equally spaced \cite{kirk2005volvox,matt2016volvox}. To represent this in our simulations, we used Spherical Centroidal Voronoi Tessellation with the package STRIPACK \cite{du2003constrained,renka1997algorithm} to distribute 1257 rotors on the surface of a sphere with a distance of approximately $2d$ between adjacent rotors, where the $200$ $\mu m$ \textit{Volvox} radius was fixed. We numerically evolved the rotor positions in time until steady states were reached. We ran a total of 30 simulations, each with random initial phases, and observed only one type of steady state, a symplectic MW. Figure~\ref{fig:UniformDynamics}(a) shows snapshots of the evolution of the MW, using a type IV Eckert projection \cite{kennedy2000understanding} to visualize the rotor phases all around the sphere in one 2D image. At $t=0$ s, the random phases are initialized; by $t=0.12$ s, patterns in the phases are beginning to form; by $t=0.3$ s, thick, solid-colored horizontal regions have formed, indicating symplectic MWs travelling from the anterior to posterior of the colony. The wave is symplectic because adjacent rotors in the direction of the posterior (the direction of the power stroke and hence the downstream direction) lag in phase behind rotors in the direction of the anterior (the direction of the recovery stroke and hence the upstream direction). In each simulation, the general shape of the symplectic MW emerged rapidly, coherent within approximately 10 beats.

Previous minimal rotor models of flagellar coordination used regularly spaced rotors, whether along a chain or a two-dimensional array \cite{brumley2012hydrodynamic, brumley2015metachronal,vilfan2006hydrodynamic,nasouri2016hydrodynamic}, which allowed for straightforward computations and comparisons of phase differences among neighboring rotors. With approximately uniform distributions of rotors on the surface of a sphere, exact linear chains of rotors do not exist. To look at the trends in rotor phases in the equatorial and meridional direction, we created chains of idealized rotors in each direction. The idealized rotors in these chains were placed at equidistant points ($\approx 2d$ apart) along the true equator and meridians (see Supplementary Material \cite{SupplementaryMaterial} for a schematic). The phase for each of the idealized rotors was computed by sampling and averaging the phases of actual rotors within a small neighborhood (radius $= 2.5d$). Let the $j^{th}$ idealized rotor in a chain have $N_j$ rotors within its local neighborhood, denoted as $\Phi_1, \Phi_2, \ldots, \Phi_{N_j}$. The phase for the $j^{th}$ idealized rotor, $\overline{\Phi}_j$, is then computed as 

\[Ae^{i\overline{\Phi}_j}=\frac{1}{N_j} \sum_{n=1}^{N_j}e^{i\Phi_n}.\]

Phases for the idealized rotor chains along meridians were computed with the same formula. Representative dynamics of idealized rotors along an equator and a single meridian are shown in Figure~\ref{fig:UniformDynamics}(b,c), respectively. These dynamics are tracked for the length of approximately two beats after the system has reached a steady state.

To better quantify the collective dynamics, we followed Wollin and Stark \cite{wollin2011metachronal} and computed the complex order parameter for the idealized rotors in the meridional and equatorial directions. Letting the neighboring phase differences of the idealized rotors be given by $\Delta \overline{\Phi}_k=\overline{\Phi}_{k+1}-\overline{\Phi}_k$ the complex order parameter is computed as
\[Ae^{i\psi}=\frac{1}{N-1} \sum_{n=1}^{N-1}e^{i\Delta \overline{\Phi}_n}.\]
\noindent A complex order parameter near 0 (A$\approx$0) means the phases are random and near 1 (A$\approx$1) means there is stable metachronism, where pairs of neighboring oscillators are phase locked with the same phase difference \cite{acebron2005kuramoto,wollin2011metachronal}. 

Along the equator we found that $A=0.99$ and $\psi=0.00$. In the meridional direction, the results were found to be dependent on the meridian chosen. To report a robust measurement, the complex order parameters were computed for 200 randomly chosen meridians. Averaging $A$ and $\psi$ across the 200 meridians considered yielded $\bar{A}=0.9969$ with a standard deviation of $0.0012$ and $\bar{\psi}=-0.1488$ with a standard deviation of $0.0093$. This strongly indicates a stable metachronal wave in the meridional direction \cite{wollin2011metachronal}. 

Overall, the steady state results share similar characteristics with previous studies of arrays of rotors above a wall \cite{brumley2015metachronal} where an average neighboring phase difference of $-0.19$ and $0$ was found along the streamwise and lateral directions, respectively, with periodic boundary conditions in the latter.

It is well known that the flagellar beat in \textit{Volvox} has some tilt out of the meridional plane estimated at 10 to 20 degrees \cite{hoops1997motility,pedley2016squirmers}. \textit{Volvox} are observed to swim with a consistent rotational spin and it is thought that this tilt in the flagellar beat causes the rotation \cite{mast1926reactions, hoops1997motility,pedley2016squirmers}. This tilt can be prescribed within the context of the present minimal model by altering the normal vector $\bfn$, which determines the plane in which the preferred orbits are situated. We proceeded by selecting $\bfn$ for each rotor to have a 15\degree~tilt from the respective meridional plane and quantifying the dynamics of the evolving rotors.

\begin{figure}
    \begin{subfigure}[t]{0.49\textwidth}
\includegraphics[width=\textwidth, trim={0cm 0cm 0cm 0cm}]{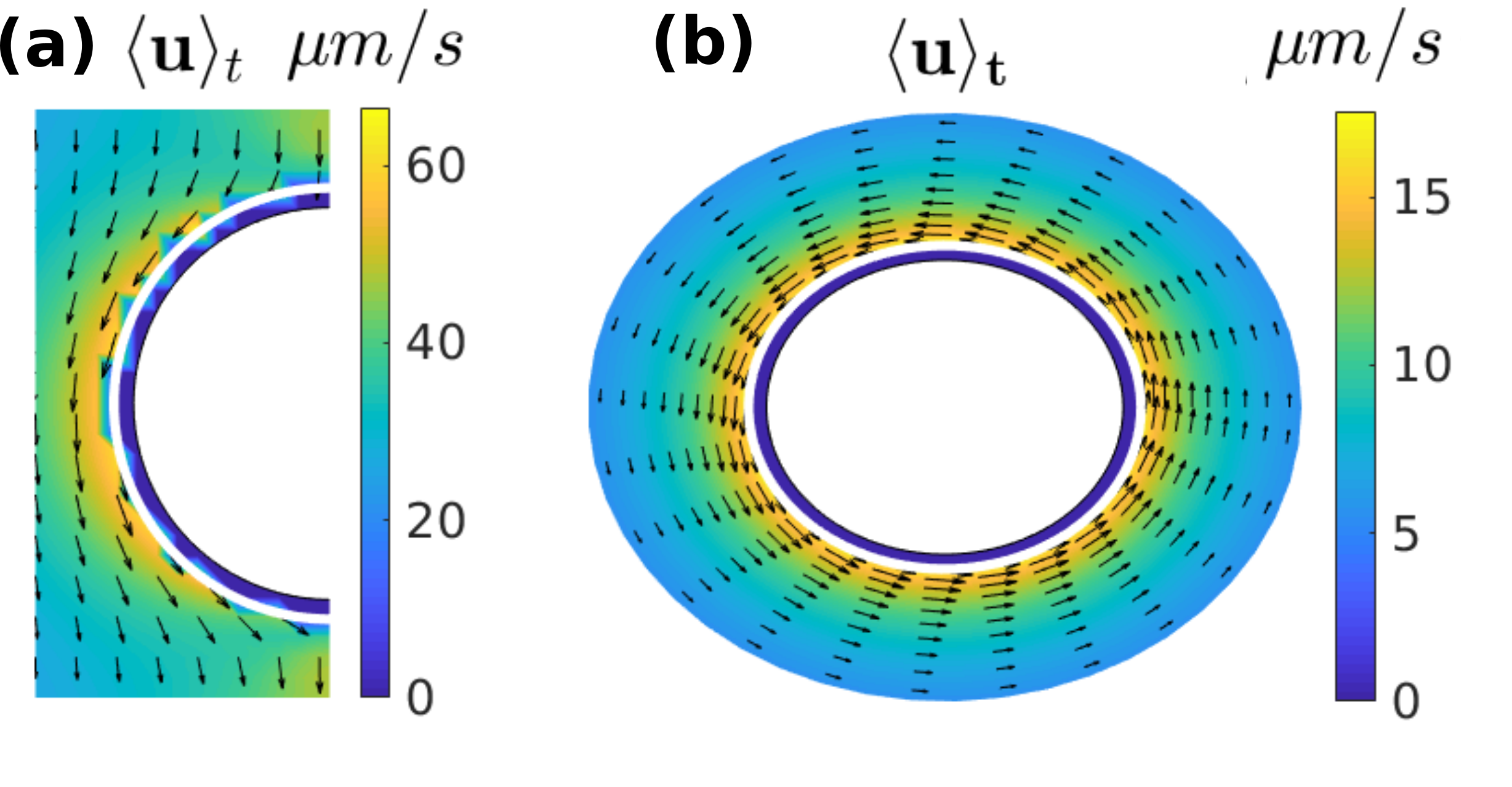}
	\end{subfigure}
	\caption{The velocity, averaged over one full periods, induced in the surrounding fluid when the rotors' preferred orbits have a tilt of $15\degree$. (a) the fluid velocity in a meridional plane. (b) the radial fluid velocity in the plane coinciding with the equator of the \textit{Volvox}. The tilt in the rotors orbits creates a clear swirling motion. In both (a) and (b) the white circle represents the \textit{Volvox} body, and the white line demarcates a distance of $20 \mu$m from the surface: velocities within this line are set to zero since far field approximations would not be applicable in this range. This figure shows the results from a horizontal steady state, it should be noted that the diagonal steady state does not have a discernibly different averaged velocity. Interestingly, this is despite the fact that the two steady states induce starkly different fluid velocities at different snapshots in time, as shown in \cite{SupplementaryMaterial}.}
	\label{fig:FluidVels}
\end{figure}

Unlike the case of no tilt, in which only a single final steady state was exhibited, simulations run with a 15\degree~tilt exhibited two possible steady states. 100 simulations with a 15\degree~tilt were run with different random initial conditions, 94 simulations reached a steady state with a horizontal simplectic metachronal wave traveling from the anterior to the posterior of the \textit{Volvox} body. This steady state is qualitatively identical to that exhibited by simulations run with no tilt, see Figure~\ref{fig:UniformDynamics}(a). This will be referred to as the horizontal steady state. Out of the 100 simulations run with random initial conditions, 6 simulations reached a steady state exhibiting  simplectic metachronal wave traveling diagonally to the anterior-posterior axis, see Figure~\ref{fig:UniformDynamics}(f). This steady state will be referred to as the diagonal steady state.

The complex order parameters in the meridional and equatorial directions were again calculated for both steady states exhibited by rotors with a 15\degree~tilt. For the horizontal steady state the complex order parameters for a chain of idealized rotors around the equator yielded $A=0.99$ and $\psi=0.00$. In the meridional direction, the complex order parameters were again averaged over 200 randomly chosen meridians yielding $\bar{A}=0.9973$ with a standard deviation of $0.019$ and $\bar{\psi}=-0.1492$ with a standard deviation of $0.0304$. This is nearly identical to the steady stated exhibited when there is no tilt. 

For the diagonal steady state, the complex order parameters for a chain of idealized rotors around the equator yielded $A=0.9994$ and $\psi=-0.1026$. In the meridional direction, the complex order parameters were again averaged over 200 randomly chosen meridians yielding $\bar{A}=0.9991$ with a standard deviation of $3.7\times 10^{-4}$ and $\bar{\psi}=-0.1335$ with a standard deviation of $0.0037$. The idealized-rotor phases for this steady state around the equator and along a meridian are shown in Figure~\ref{fig:UniformDynamics}(d)-(e).

 Prescribing a tilt out of the meridional plane to the rotors' preferred orbits was motivated by the question of whether this is indeed the origin of the swirling motion exhibited by \textit{Volvox}. While the current model does not incorporate movement of the \textit{Volvox} body, we can examine the velocity induced in the surrounding fluid. Just as the influence one rotor has on another can be estimated by using the far-field approximation of a Stokeslet, the fluid flow away from the \textit{Volvox} body can be estimated by summing the far-field approximations of the fluid flow induced by each rotor. Figure~\ref{fig:FluidVels} shows the far-field fluid flow induced by the rotors with a 15\degree~tilt when horizontal steady state is reached. The time-averaged velocity magnitude (color) and direction (black vector) are shown in the meridional plane in Figure~\ref{fig:FluidVels}(a) and in the plane coinciding with the equator in Figure~\ref{fig:FluidVels}(b). The far-field approximation is demarcated by the white line at a distance of $20 \mu$m from the \textit{Volvox} body and velocities within this line are set to 0 since the far-field approximations used are not applicable within this range. 
 
 As seen in Figure~\ref{fig:FluidVels}(b), a clear rotational velocity is observed in the surrounding fluid when a tilt is induced in the rotor orbit. The spatial distribution of the time-averaged velocities in a meridional plane also compares well with previous laboratory measurements of in-vivo \textit{Volvox} \cite{brumley2015metachronal}, however it should be noted the magnitudes are roughly 3 times as small. In previous studies, magnitudes of Stokeslets were fit to experimental data from a single \textit{Volvox} somatic cell and its flagella \cite{brumley2014flagellar}; the fit values were approximately three times the forces exerted in the present model. Since velocity scales linearly with force in Stokes flow, if the present forces were scaled by a factor of 3, the fluid velocity magnitudes would match the previous experimental data very well.

Previous biological studies of \textit{Volvox} have established that there is clear large-scale flagellar coordination across the algae body \cite{brumley2012hydrodynamic, brumley2015metachronal}. Minimal models of the flagella coupling have thus far only assessed coordination above a planar surface \cite{brumley2012hydrodynamic, brumley2015metachronal, vilfan2006hydrodynamic}, or a linear chain of rotors outside a spherical surface all beating in one direction \cite{nasouri2016hydrodynamic}. The present model considers a biologically-relevant distribution of rotors exterior to a sphere and reproduces the experimentally-observed flagellar coordination of \textit{Volvox}. We note that the qualtitative coordination obtained with this model does not differ significantly from studies above a planar surface, suggesting that an array of rotors above a plane well-approximates the coupling of a patch of rotors above a sphere. However, considering a distribution of rotors around a sphere allows the generation of more pertinent flows to the actual organism. For example, prescribing a tilt to the preferred rotor orbit to mimic the tilt of \textit{Volvox} flagella generates a swirl in the flow around the spherical body that can not be captured with a planar geometry. To our knowledge, there have been no published experimental results showing the different horizontal and diagonal steady states that our model revealed; we hope that our study will inspire more experiments to investigate such behavior. 

 The present study has not addressed the sensitivity of the steady state synchronization to such parameters as the spacing between rotors, the stiffness of the springs determining the rotors preferred path, and the rotor tilt. This will be explored in subsequent studies. Additionally, the organized flagellar beat exhibited across the surface of \textit{Volvox} allows the organism to undergo phototaxis. An improvement of the present model would be to incorporate the swimming of the organism. Our study also lays the groundwork for future investigations of spatially-varying densities of somatic cells, varying flagellar beat forms due to light, and phototaxis \cite{ueki20105000}.


%

\end{document}